\documentclass[twocolumn,preprintnumbers,amsmath,amssymb]{revtex4}

\usepackage{graphicx}
\usepackage{dcolumn}
\usepackage{bm}
\usepackage{graphics}
\usepackage[final]{hyperref}
\usepackage{amssymb}
\usepackage{textcomp}
\usepackage{scrpage2}
\begin{document}
\begin{minipage}[b][50pt][b]{1pt}
\includegraphics[width=240pt]{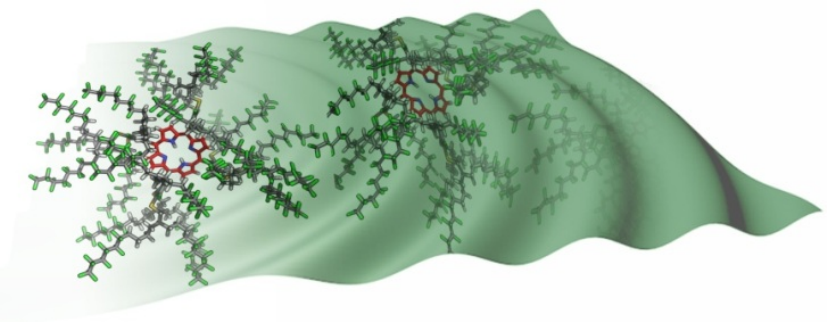}
\end{minipage}
\preprint{PCCP, DOI: 10.1039/C3CP51500A}
\date{}
\title{Matter-wave interference with particles selected from \\a molecular library
with masses exceeding 10 000 amu}

\author{Sandra Eibenberger}\affiliation{University of Vienna, Faculty of Physics, VCQ, QuNaBioS,
Boltzmanngasse 5, 1090 Vienna (Austria)}
\author{Stefan Gerlich}\affiliation{University of Vienna, Faculty of Physics, VCQ, QuNaBioS,
Boltzmanngasse 5, 1090 Vienna (Austria)}
\author{Markus Arndt}\email{markus.arndt@univie.ac.at}\affiliation{University of Vienna, Faculty of Physics, VCQ, QuNaBioS,
Boltzmanngasse 5, 1090 Vienna (Austria)}
\author{Marcel Mayor}\email{marcel.mayor@unibas.ch}\affiliation{Department of Chemistry, University of Basel, St. Johannsring 19,
4056 Basel (Switzerland),\\Karlsruhe Institute of Technology (KIT), Institute of Nanotechnology, 
P.O. Box 3640, 76021 Karlsruhe (Germany)}
\author{Jens T\"uxen}\affiliation{Department of Chemistry, University of Basel, St. Johannsring 19,
4056 Basel (Switzerland)}

%

\date{\today}

\begin{abstract}
The quantum superposition principle, a key distinction between quantum physics and classical mechanics, is often perceived as a philosophical challenge to our concepts of reality, locality or space-time since it contrasts our intuitive expectations with experimental observations on isolated quantum systems. While we are used to associating the notion of localization with massive bodies, quantum physics teaches us that every individual object is associated with a wave function that may eventually delocalize by far more than the body's own extension. Numerous experiments have verified this concept at the microscopic scale but intuition wavers when it comes to delocalization experiments with complex objects. While quantum science is the uncontested ideal of a physics theory, one may ask if the superposition principle can persist on all complexity scales. This motivates matter-wave diffraction and interference studies with large compounds in a three-grating interferometer configuration which also necessitates the preparation of high-mass nanoparticle beams at low velocities. Here we demonstrate how synthetic chemistry allows us to prepare libraries of fluorous porphyrins which can be tailored to exhibit high mass, good thermal stability and relatively low polarizability, which allows us to form slow thermal beams of these high-mass compounds, which can be detected in electron ionization mass spectrometry. We present successful superposition experiments with selected species from these molecular libraries in a quantum interferometer, which utilizes the diffraction of matter waves at an optical phase grating. We observe high-contrast quantum fringe patterns with molecules exceeding a mass of 10\,000 amu and 810\,atoms in a single particle.
\end{abstract}

\maketitle

\section{\label{sec:Intro}Introduction}
Quantum physics has long been regarded as the science of 'small things', but experimental progress throughout the last two decades has led to the insight that it can also be observable for mesoscopic or even macroscopic objects. This applies for instance to the superposition of macroscopic numbers of electrons in superconducting quantum devices \cite{Mooij1999}, the realization of large quantum degenerate atomic clouds in Bose-Einstein condensates \cite{Anderson1995}, the cooling of micromechanical oscillators to their mechanical ground state. 

Quantum superposition studies with complex molecules \cite{Arndt1999} became possible with the advent of new matter wave interferometers \cite{Brezger2003,Gerlich2007,Haslinger2013} and techniques for slow macromolecular beams \cite{Deachapunya2008}. These interferometers were for instance practically used to enable quantum enhanced measurements of internal molecular properties. The quantum fringe shift of a molecular interference pattern in the presence of external electric fields resulted in information for example on electric polarizabilities \cite{Berninger2007}, dipole moments \cite{Eibenberger2011}, or configuration changes \cite{Gring2010}. Molecule interferometry can complement mass spectrometry \cite{Gerlich2008} and help to distinguish constitutional isomers \cite{Tuxen2010}. In addition to their applications in chemistry, quantum interference experiments with massive molecules currently set the strongest bound on certain models that challenge the linearity of quantum mechanics \cite{Bassi2013}.

Further exploration of the frontiers of de Broglie coherence now profits from new capabilities in tailoring molecular properties to the needs of quantum optics. Our quantum experiment dictates the design of the molecules and the challenges increase with the number of atoms involved. In order to realize a molecular beam of sufficient intensity, the model compounds must be volatile, thermally stable and accessible in quantities of several hundred milligrams. Moreover, in order to minimize absorption at the wavelength of the optical diffraction (see below) they need to feature low absorption but sufficient polarizability at 532\,nm. In reply to these needs, a dendritic library concept is particularly appealing since it can be scaled up to complex particles, once a suitable candidate has been identified.

To meet these requirements we have functionalized organic chromophores with extended perfluoroalkyl chains. Such compounds show low inter-molecular binding and relatively high vapor pressures \cite{Stock2004,Krusic2005}. They possess strong intra-molecular bonds and therefore sufficient thermal stability. In addition we start with a porphyrin core which is compatible with the required optical and electronic properties \cite{Meot-Ner1973}. In the past, monodisperse fluorous porphyrins were generated by substituting the four \textit{para}-fluorine substituents of tetrakis pentafluorophenylporphyrin (\textbf{TPPF20}) by dendritically branched fluorous moieties \cite{Tuxen2011}. With this approach, molecules composed of 430\,atoms were successfully synthesized and applied in quantum interference experiments \cite{Gerlich2011}.

With increasing complexity it becomes more challenging to purify monodisperse particles in sufficient amounts. Here we profit from the fact that our interferometer arrangement allows us to work with compound mixtures since each molecule interferes only with itself. By substituting some of the twenty fluorine atoms of \textbf{TPPF20} with a branched, terminally perfluorinated alkylthiol (\textbf{1}), we obtain a mixture of compounds with molecular masses that differ exactly by an integer multiple of a particular value as molecular library. The molecular beam density is sufficiently low for the molecules not to interact with each other and the individual library compounds can be mass-specifically detected in a quadrupole mass spectrometer (QMS Extrel, 16\,000\,amu).

Our synthetic approach is based on the fact that pentafluorophenyl moieties can be used to attach up to five polyfluoroalkyl substituents in nucleophilic aromatic substitution reactions. Substitutions at \textbf{TPPF20} with its 20 potentially reactive fluorine substituents lead to a molecular library of derivatives with a varying number of fluorous side chains.

\section{Results and Discussion}
We used sodium hydride as a base, microwave radiation as a heating source and diethylene glycol dimethyl ether (diglyme) as fluorophilic solvent. \textbf{TPPF20}, and a large excess of the thiol \textbf{1} (60 equivalents) and sodium hydride in diglyme were heated in a sealed microwave vial to 220\,\textdegree C for 5\,minutes.\footnote{
\textbf{Synthetic protocol and analytical data of the porphyrin libraries \textit{L}}:
General Remarks: All commercially available starting materials were of reagent grade and used as received. Microwave reactions were carried out in an \textit{Initiator 8} (400\,W) from \textit{Biotage}. Glass coated magnetic stirring bars were used during the reactions. The solvents for the extractions were of technical grade and distilled prior to use. Matrix Assisted Laser Desorption Ionization Time of Flight (MALDI-ToF) mass spectra were performed on an \textit{Applied Bio Systems Voyager-De\rm{\texttrademark} Pro} mass spectrometer or a \textit{Bruker microflex} mass spectrometer. Significant signals are given in mass units per charge (m/z) and the relative intensities are given in brackets.\\
\textbf{Porphyrin library \textit{L}:}
The thiol \textbf{1} was synthesized in seven reaction steps in an overall yield of 70\% as reported elsewhere \cite{Tuxen2011}. 5,10,15,20-Tetrakis(pentafluoro-phenyl)-porphyrin (\textbf{TPPF20}, 4.0\,mg, 4.10\,$\rm{mu}$mol, 1.0\,eq.), thiol \textbf{1} (193\,mg, 246\,$\rm{mu}$mol, 60\,eq.) and sodium hydride (60\% dispersion in mineral oil, 14.8\,mg, 369\,$\rm{mu}$mol, 90\,eq.) were added to diglyme (4\,mL) in a microwave vial. The sealed tube was heated under microwave irradiation to 220\,\textdegree C for 5 minutes. After cooling to room temperature the reaction mixture was quenched with water and subsequently extracted with diethylether. The organic layer was washed with brine and water, dried over sodium sulfate and evaporated to dryness. The resulting product mixture (183\,mg) was analyzed by MALDI-ToF mass spectrometry. \textbf{MS} (MALDI-ToF, m/z): 12\,403 (29\%), 11\,645 (52\%), 10\,884 (100\%), 10\,121 (78\%), 9\,339 (15\%), 8\,597 (7\%).} 
After aqueous workup the resulting mixture was analyzed by MALDI-ToF mass spectrometry (Figure 1) and subsequently used in our quantum interference experiments without further purification. We find up to 15 substituted fluorous thiol chains reaching to a molecular weight well beyond 10\,000 amu.

\begin{figure}
\includegraphics[width=\columnwidth]{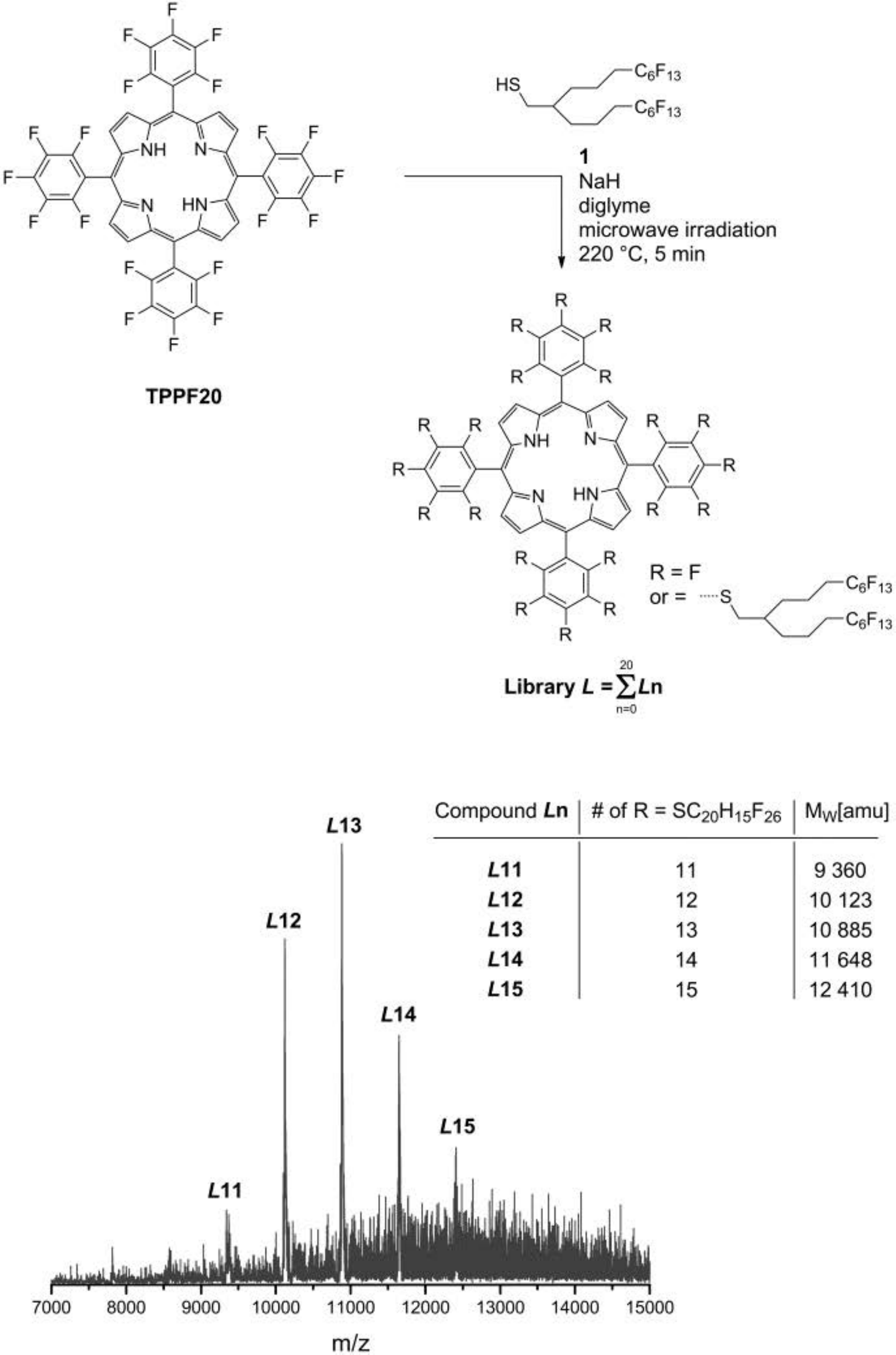}
\caption{\label{fig:1}Synthetic scheme and MALDI-ToF mass spectrum of the fluorous porphyrin library \textbf{textit{L}}. High-mass matter-wave experiments were performed with component \textbf{\textit{L}}\textbf{12} of the library \textbf{\textit{L}}. This structure is composed of 810 atoms and has a nominal molecular weight of 10\,123 amu.}
\end{figure}

\begin{figure*}
\includegraphics[width=\textwidth]{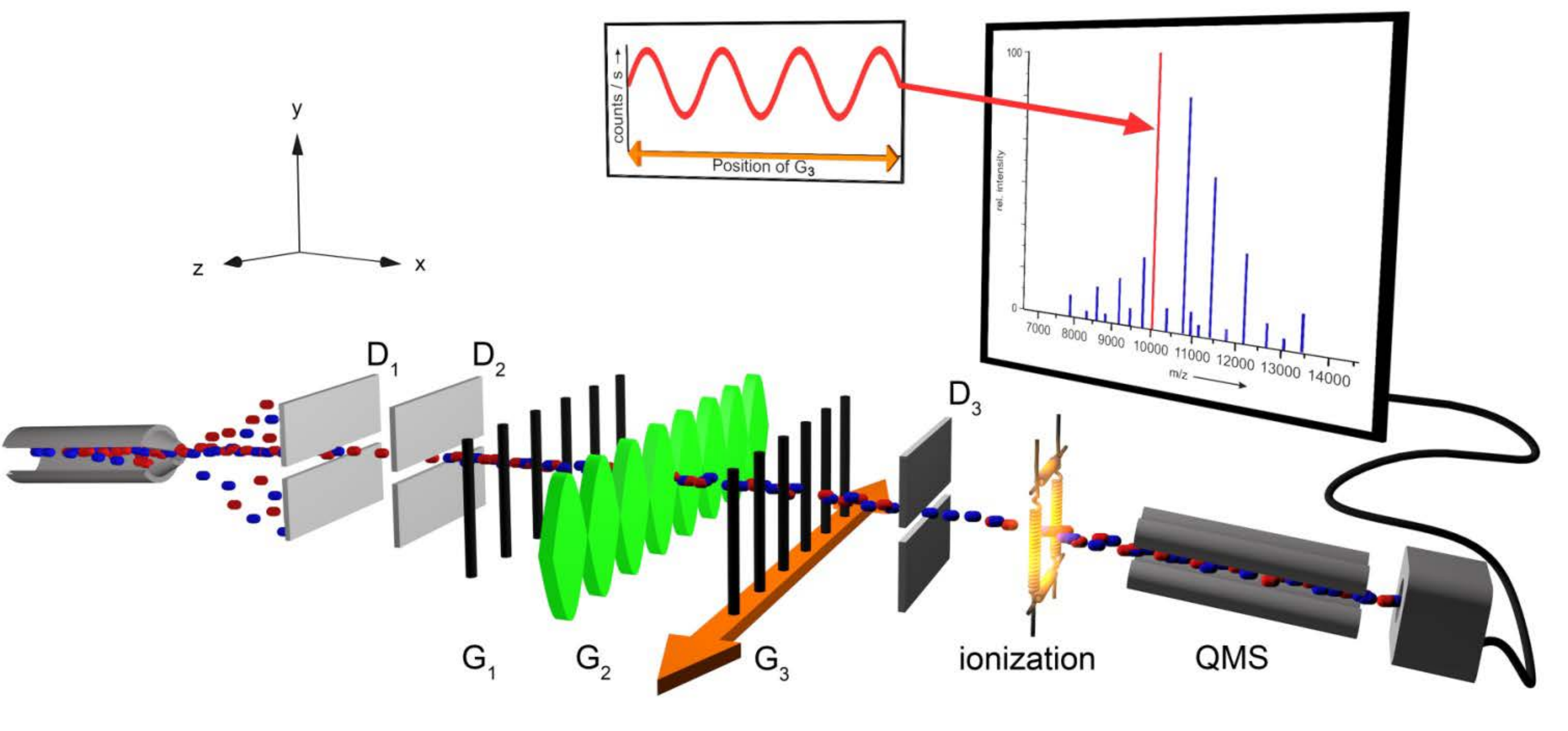}
\caption{\label{fig:2}KDTL interferometer setup: The molecules are evaporated in a furnace. Three height delimiters, D$_{\rm{1}}$ - D$_{\rm{1}}$, define the particle velocity by selecting a flight parabola in the gravitational field. The interferometer consists of three gratings with identical periods of $d$ = 266\,nm. G$_{\rm{1}}$ and G$_{\rm{3}}$ are SiN$_{\rm{x}}$ gratings, whereas G$_{\rm{2}}$ is a standing light wave which is produced by retro-reflection of a green laser at a plane mirror. A phase modulation $\Phi\propto(\alpha\cdot P)/(v\cdot\omega_y)$ is imprinted onto the molecular matter wave via the optical dipole force which is exerted by the light grating onto the molecular optical polarizability $\alpha_{opt}$. Here $P$ is the laser power, $v$ the molecular velocity, $\omega_x\cong\rm{18\,\mu m}$ and $\omega_y\cong\rm{945\,\mu m}$ the Gaussian laser beam waists. The transmitted molecules are detected using electron ionization quadrupole mass spectrometry after their passage through G$_{\rm{3}}$, which can be shifted along the z-axis to sample the interference pattern.}
\end{figure*}

In order to study the delocalized quantum wave nature of compounds in the fluorous library we utilize a \textit{Kapitza-Dirac-Talbot-Lau} interferometer (KDTLI), which has already proven to be a viable tool with good mass scalability in earlier studies \cite{Gerlich2007,Gerlich2011}. The interferometer is sketched in Figure \ref{fig:2}: a molecular beam is created by thermal evaporation of the entire library in a Knudsen cell. The mixture traverses three gratings G$_{\rm{1}}$, G$_{\rm{2}}$, and G$_{\rm{3}}$, which all have the same period of d $\cong$ 266\,nm. The molecules first pass the transmission grating G$_{\rm{1}}$, a SiN$_{\rm{x}}$ mask with a slit opening of $s\approx$ 110\,nm, where each molecule is spatially confined to impose the required spatial coherence by virtue of \textit{Heisenberg}'s uncertainty principle \cite{Nairz2002}. This is sufficient for the emerging quantum wavelets to cover several nodes of the optical phase grating G$_{\rm{2}}$, 10.5\,cm further downstream. The standing light wave G$_{\rm{2}}$ is produced by retro-reflection of a green laser ($\lambda_{\rm{L}}$= 532 nm) at a plane mirror.

When the molecular matter wave traverses the standing light wave, the dipole interaction between the electric field of power $P$ and the molecular optical polarizability $\alpha_{\rm{opt}}$ entails a periodic phase modulation $\Phi=\Phi_0\cdot\sin^2{(2\pi z/\lambda_L)}$ with the maximum phase shift $\Phi_0=8\sqrt{2\pi}\alpha_{opt}P/(\hbar c\omega_yv_x)$. Here $v_x$ is the forward directed molecular velocity, $\omega_y\cong$ 945\,$\rm{\mu}$m is the \textit{Gaussian} laser beam waist along the grating slits and $z$ is the coordinate along the laser beam.

Interference of the molecular wavelets behind G$_{\rm{2}}$ leads to a molecular density pattern of the same period $d$ in front of the third grating. As long as we can neglect photon absorption by the molecules in the standing light wave G$_{\rm{2}}$ acts effectively as a pure phase grating. The expected fringe visibility $V$ is then given by \cite{Hornberger2009} $V=2(\sin(\pi f)/\pi f)^2J_2(-sgn(\Phi_0\sin(\pi L/L_T))\Phi_0\sin(\pi L/L_T))$. Here $f$ designates the grating open fraction, i.e. the ratio between the open slit width $s$ and the grating period $d$. $J_2$ is the Bessel function of second order, $L$ the separation of the gratings, $L_T=d^2/\lambda_{dB}$ the Talbot length and $\lambda_{dB}=h/(mv)$ the \textit{de Broglie} wavelength with $h$ as \textit{Planck}'s quantum of action, $m$ the molecular mass and $v$ the modulus of its velocity.

G$_{\rm{3}}$ is again a SiN$_{\rm{x}}$ structure and lends spatial resolution to the detector. The interferogram is sampled by tracing the transmitted particle beam intensity as a function of the lateral ($z$) position of G$_{\rm{3}}$ and the mass selection is performed in the mass spectrometer behind this stage.

\textit{Talbot-Lau} interferometers \cite{Clauser1997} offer the important advantage over simple grating diffraction that the required grating period $d$ only weakly depends on the molecular \textit{de Broglie} wavelength: $d\propto\sqrt{\lambda_{dB}}$. The setup accepts a wide range of velocities and low initial spatial coherence. This facilitates the use of dilute thermal molecular beams. Diffraction at the standing light wave G$_{\rm{2}}$ avoids the dephasing caused by the \textit{van der Waals} interaction between the molecules and a dielectric wall. This is indeed present in G$_{\rm{1}}$ and G$_{\rm{3}}$ but can be neglected there since the molecular momentum distribution at G$_{\rm{1}}$ is wider than that caused by the grating and any phase shift in G$_{\rm{3}}$ is irrelevant if we are only interested in counting the particles that reach the mass spectrometer.

Indistinguishability in all degrees of freedom is the basis for quantum interference \cite{Dirac1958} and naturally given if a single molecule interferes only with itself. We only have to make sure that every molecule contributes to the final pattern in a similar way, which is true for all members of the library at about the same mass, independent of their internal state.

Differences between various molecules, such as their isotopic distribution or the addition of a single atom are still acceptable. The KDTLI can tolerate a wavelength distribution of $\Delta\lambda_{dB}/\lambda_{dB}\leq\rm{20\,\%}$ and still produce a quantum fringe visibility in excess of the classical threshold.

We here present quantum interference collected at the mass of one specific library compound, particularly for \textbf{\textit{L}}\textbf{12}$=\rm{C_{284}H_{190}F_{320}N_4S_{12}}$ which has 12 fluorous side chains, a mass of 10\,123\,amu and 810 atoms bound in a single hot nanoparticle.

All molecules of the library were evaporated at a temperature of about 600\,K. We selected the velocity class around $v$ = 85\,m/s ($\Delta v_{\rm{FWHM}}$ = 30\,m/s) – corresponding to a most probable \textit{de Broglie} wavelength of approximately 500\,fm. This is about four orders of magnitude smaller than the diameter of each individual molecule. We detected the signal by electron ionization quadrupole mass spectrometry. During the interference measurements the mass filter was set to the target mass of \textbf{\textit{L}}\rm{\textbf{12} and only this compound contributed to the collected interference pattern.

The molecular beam was dilute enough to prevent classical interactions between any two molecules within the interferometer. Given that 80\,mg of library \textbf{\textit{L}} molecules were evaporated in 45 minutes, we estimate a flux at the source exit of ~$2\times10^{15}$ particles per second. Including the acceptance angle of the instrument, the velocity selection as well as the grating transmission we estimate a molecular density inside the interferometer of 30\,mm$^{\rm{-3}}$. This corresponds to a mean particle distance of about 300\,$\rm{\mu}$m which is sufficient to exclude interactions with other neutral molecules in the beam.

The average flight time of a molecule through the tightly focused standing light wave amounts to about 400\,ns, i.e. much longer than the time scale of molecular vibrations (10$^{\rm{-14}}$ - 10$^{\rm{-12}}$\,s) and rotations (~10$^{\rm{-10}}$\,s). Therefore, the mean scalar polarizability governs the interaction with the standing light wave although the optical polarizability is generally described by a tensor. Thermal averaging also occurs for the orientation of any possibly existing molecular electric dipole moment \cite{Eibenberger2011}. The internal molecular states are decoupled from de Broglie interference as long as we exclude effects of collisional or thermal decoherence \cite{Hackermuller2003,Hackermuller2004} or external force fields \cite{Berninger2007}.

The thermal mixture of internal states is another reason why two-particle interference, i.e. mutual coherence of two macromolecules, is excluded in our experiments. The chances of finding two of them in the same indistinguishable set of all internal states – electronic, vibrational and rotational levels, configuration, orientation and spin – is vanishingly small.

\begin{figure}
\includegraphics[width=\columnwidth]{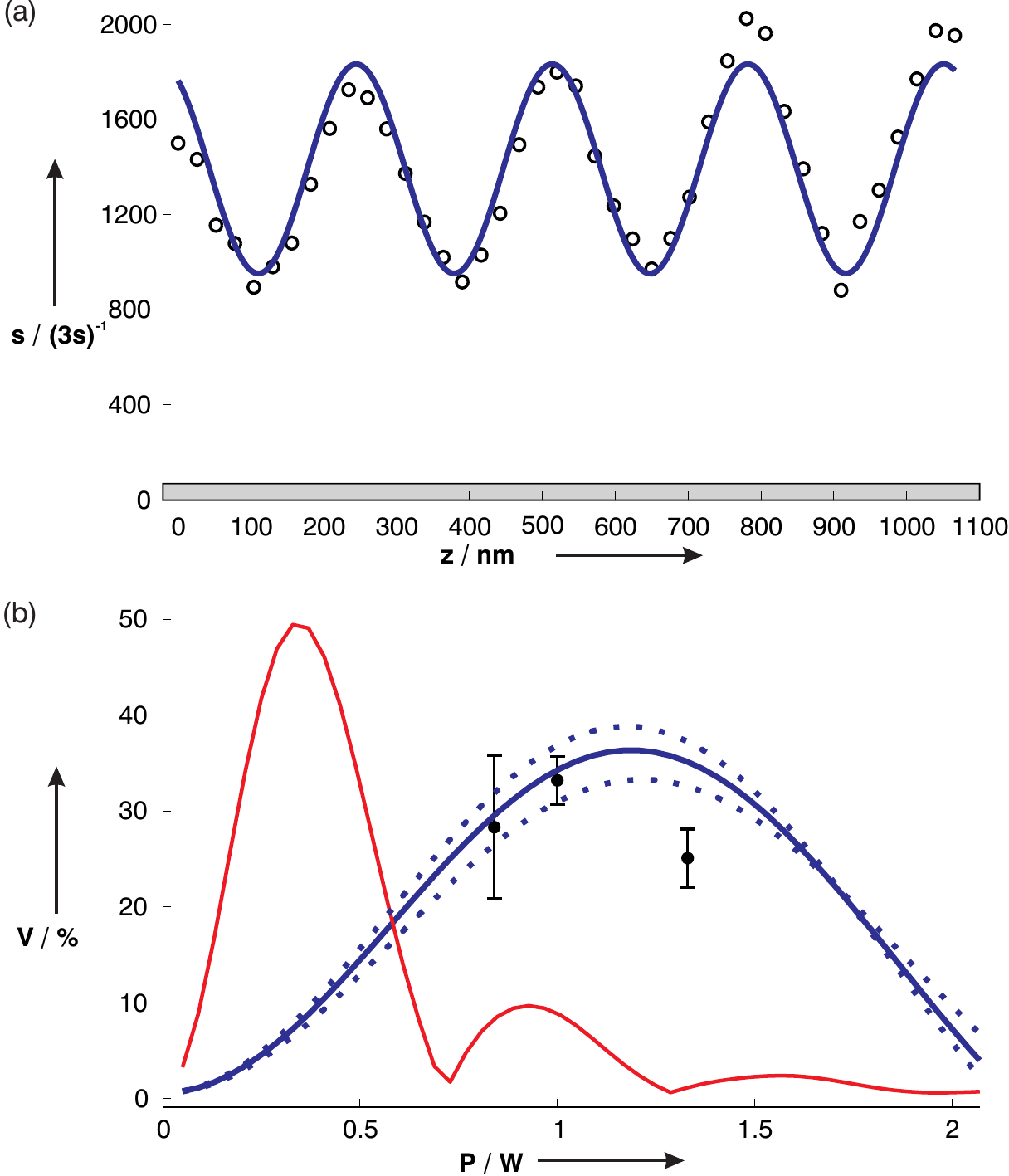}
\caption{\label{fig:3}(a) Quantum interference pattern of \textbf{\textit{L}}\textbf{12} recorded at a laser power of $P\cong\rm{1\,W}$. The circles represent the experimental signal $s$ as a function of the position $z$ of the third grating. The solid line is a sinusoidal fit to the data, with a quantum fringe visibility of $V$ = 33(2)\,\%. The shaded area represents the background signal of the detector. A classical picture predicts a visibility of only 8\% for the same experimental parameters. b) Measured fringe visibility $V$ as a function of the diffracting laser power $P$. The expected contrast according to the quantum and the classical model are plotted as the blue and red lines, respectively \cite{Hornberger2009}. The dashed blue lines correspond to the expected quantum contrast when the mean velocity is increased (reduced) by 5\,ms$^{\rm{-1}}$.}
\end{figure}

In Figure \ref{fig:3}(a) we show a high contrast quantum interference pattern of \textbf{\textit{L}}\textbf{12}. In contrast to far-field diffraction where the fringe separation is governed by the de Broglie wave-length of the transmitted molecules,\cite{Nairz2003,Juffmann2012} near-field interferometry of the \textit{Talbot-Lau} type generates fringes of a fixed period, which are determined by the experimental geometry. Specifically, the expected interference figure in our KDTLI configuration is a sine curve whose contrast varies with the phase-shifting laser power as well as with the molecular beam velocity and polarizability.
We distinguish the genuine quantum interferogram \cite{Brezger2003,Hornberger2009} from a classical shadow image by comparing the expected and experimental interference fringe visibility (contrast) with a classical model.

The far off-resonance optical polarizability is assumed to be well approximated by the static value $\alpha_{opt}\cong\alpha_{stat}\cong\rm{410}\rm{\AA^3}\times\rm{4}\pi\epsilon_0$ as estimated using Gaussian G09 \cite{Frisch2009} with the 6-31 G basis set. The absorption cross section of \textbf{\textit{L}}\textbf{12} at 532\,nm was estimated using the value of pure tetraphenylporphyrine dissolved in toluene \cite{Dixon2005} assuming that the perfluoroalkyl chains contribute at least an order of magnitude less to that value \cite{Gotsche2007}. We thus find $\sigma_{532}\cong 1.7\times\rm{10^{-21}\,m^2}$.

In Figure \ref{fig:3}(b) we show the expected classical and quantum contrast as a function of the diffracting laser power. Our experimental contrast is derived from the recorded signal curves, such as shown in Figure \ref{fig:3}(a) by $V=(S_{max}-S_{min})/(S_{max}+S_{min})$, where $S_{max}$ and $S_{min}$ are the maxima and minima of the sine curve fitted to the data. The contrast it is plotted as the black dots in Figure \ref{fig:3}(b). These points are in good agreement with the quantum prediction (blue line). The classical effect, describing the shadow image by two material gratings and one array of dipole force lenses \cite{Hornberger2009} is shown as the red line.

The experiment clearly excludes this classical picture. Since the amount of precious molecular material and thermal degradation processes in the source limited the measurement time we performed our experiments only in the parameter regime that maximizes the fringe visibility, to show the clear difference between the quantum and classical contrast. The experimental fringe visibility reproduces well the maximally expected quantum contrast.

\section{Conclusion}
We have shown that a library approach towards stable and volatile high-mass molecules can substantially extend the complexity range of molecular \textit{de Broglie} coherence experiments to masses in excess of 10\,000 amu. Our data confirm the fully coherent quantum delocalization of single compounds composed of about 5\,000 protons, 5\,000 neutrons and 5\,000 electrons. The internal complexity, number of vibrational modes and also the internal energy of each of these particles is higher than in any other matter-wave experiment so far.

\begin{acknowledgments}
This work was supported by the FWF programs Wittgenstein (Z149-N16) and CoQuS (W1210-2), the European Commission in the project NANOQUESTFIT (304 886), the Swiss National Science Foundation, the NCCR 'Nanoscale Science', and the Swiss Nanoscience Institute (SNI). The authors thank Michel Rickhaus for the artwork of figure 2.
\end{acknowledgments}
\bibliography{references}

\end{document}